\documentclass{aa}

\usepackage[varg]{txfonts}
\usepackage{graphicx}
\usepackage{amsmath}
\usepackage[colorlinks=true,linkcolor=magenta,citecolor=blue,filecolor=cyan,urlcolor=red,final=true]{hyperref}

\makeatletter
\renewcommand*\aa@pageof{, page \thepage{} of \pageref*{LastPage}}
\makeatother

\begin{document} 

   \title{A type II solar radio burst without a coronal mass ejection}
	\titlerunning{A type II solar radio burst without a coronal mass ejection}

   \author{D.~E.~Morosan \inst{1}
           \and
        J. Pomoell \inst{1}
        \and
        A.~Kumari \inst{1}
        \and
        E.~K.~J.~Kilpua \inst{1} 
        \and
        R. Vainio \inst{2}
        }

   \institute{Department of Physics, University of Helsinki, P.O. Box 64, FI-00014 Helsinki, Finland \\
              \email{diana.morosan@helsinki.fi}
        \and
             Department of Physics and Astronomy, University of Turku, 20014, Turku, Finland
             }

   \date{Received ; accepted }

 
  \abstract
    {The Sun produces the most powerful explosions in the solar system, solar flares, that can also be accompanied by large eruptions of magnetised plasma, coronal mass ejections (CMEs). These processes can accelerate electron beams up to relativistic energies through magnetic reconnection processes during solar flares and CME-driven shocks. Energetic electron beams can in turn generate radio bursts through the plasma emission mechanism. CME shocks, in particular, are usually associated with type II solar radio bursts. }
    {However, on a few occasions, type II bursts have been reported to occur either in the absence of CMEs or shown to be more likely related with the flaring process. It is currently an open question how a shock generating type II bursts forms without the occurrence of a CME eruption. Here, we aim to determine the physical mechanism responsible for a type II burst which occurs in the absence a CME. }
    {By using radio imaging from the Nan{\c c}ay Radioheliograph, combined with observations from the Solar Dynamics Observatory and the Solar Terrestrial Relations Observatory spacecraft, we investigate the origin of a type II radio burst that appears to have no temporal association with a white-light CME. }
    {We identify a typical type II radio burst with band-split structure that is associated with a C-class solar flare. The type II burst source is located above the flaring active region and ahead of disturbed coronal loops observed in extreme ultraviolet images. The type II is also preceded by type III radio bursts, some of which are in fact J-bursts indicating that accelerated electron beams do not all escape along open field lines. The type II sources show single-frequency movement towards the flaring active region. The type II is located ahead of a faint extreme-ultraviolet (EUV) front propagating through the corona. }
    {Since there is no CME detection, a shock wave is most likely generated by the flaring process or the bulk plasma motions associated with a failed eruption. The EUV front observed is likely a freely propagating wave that expands into surrounding regions. The EUV front propagates at an initial speed of approximately 450~km/s and it is likely to steepen into a shock wave in a region of low Alfv\'en speed as determined from magneto-hydrodynamic modelling of the corona.    }

   \keywords{Sun: corona -- Sun: radio radiation -- Sun: particle emission -- Sun: coronal mass ejections (CMEs) -- Sun: activity -- Sun: flares}

\maketitle


\section{Introduction}

{Solar radio bursts can be produced by energetic processes in the solar corona such as solar flares and coronal mass ejections (CMEs). These processes accelerate electron beams that can in turn generate emission at radio wavelengths through the plasma emission mechanism \cite[see e.g.,][and the references therein]{kl02}. Solar radio bursts are classified based on their shape and characteristics in dynamic spectra as type I--V \citep[][]{wi63}. Type II and type IV bursts are usually associated with a combination of a CME and flare or a CME only, but rarely with solar flares only \citep[e.g.][]{ magdalenic12, su2015, pankaj2016}, which suggest that a CME is critical in producing them \citep[e.g.][]{smerd1970, du76, ge86, mo19b, salas2020polarisation, mo21}. Type III bursts in turn are associated with flares or other activity on the Sun in the absence of flaring events and they represent signatures of electron beams travelling along open magnetic field lines \citep[e.g.][]{re14}.  }

{Type II radio bursts, in particular, are usually associated with electron beams accelerated by CME-driven shock waves \citep[][]{smerd1970, Gopalswamy2005, kumari2019}. Type II bursts consist of slowly drifting emission lanes observed at metric to decametric wavelengths in dynamic spectra at the fundamental and/or harmonic of the plasma frequency \citep{ma96,ne85}. The fundamental and harmonic emission lanes can often show split bands \citep[e.g.,][]{Smerd1975, vr01,vr02,ku2017b} and numerous fine structures composing these lanes \citep[e.g.,][]{ma20}. For example, fine-structured bursts called `herringbones' are characterised by short-duration drifting bursts towards either low or high frequencies stemming from a type II `backbone' \citep{ho83,ca87,ca89}, but sometimes also occurring without a type II backbone \citep[][]{ho83,mo19a}. There is also a wealth of other  fine  structures composing the  emission  lanes  of  type  II  bursts, recently documented by \citet[][]{ma20}. Type II and associated fine structures are usually classified as the radio signatures of expanding CME-driven shocks in the solar corona \citep[e.g.,][]{st74, ma12, liu09, zu18, ca13, kumari2017a, mo19a}. }

\begin{figure*}[ht]
    \centering
    \includegraphics[width=0.82\linewidth]{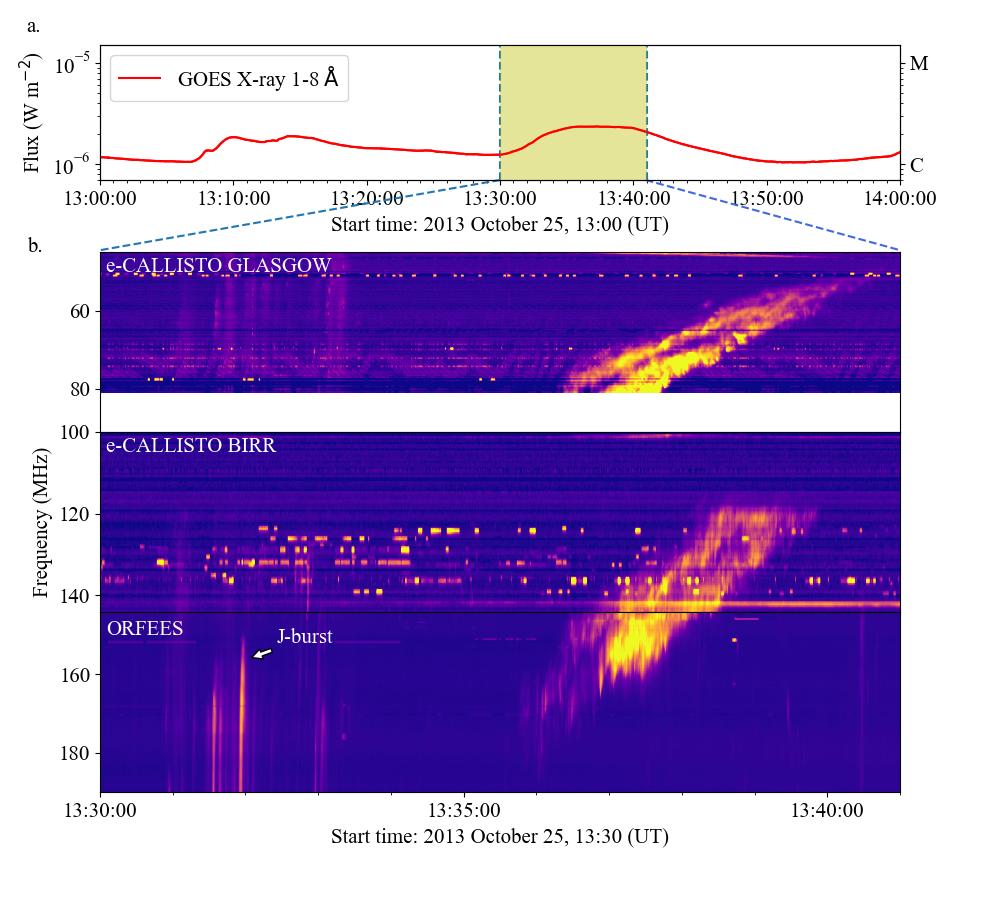}
    \caption{Flare X-ray time series together with a dynamic spectrum of a type II radio burst. (a) GOES X-Ray light curve in the 1-8~$\AA$ channel. A C-class flare associated with the Type II burst is shown inside the yellow region. (b) Combined dynamic spectrum from e-CALLISTO Glasgow (40--80-~MHz), e-CALLISTO Birr (100--145~MHz) and ORFEES (145--200~MHz) during the time period of the yellow shaded region in (a). The dynamic spectrum shows Type III radio bursts followed by a type II bursts with fundamental and harmonic lanes of emission. }
    \label{fig:fig1}
\end{figure*}

{Some studies, however, suggest that not all type IIs are associated with CME shocks \citep[e.g.,][]{ magdalenic12, su2015, pankaj2016}. A different mechanism has been suggested for high-frequency type II bursts ($>$~150~MHz), that is, flare generated shock waves, due to some type II speeds being much larger than that of the associated CME \citep[e.g.,][]{pohjolainen2008, magdalenic08, magdalenic10}. A recent study of a type II that occured at low frequencies, <100~MHz, suggested that its origin was a piston-driven shock wave originating from the active region jet during the eruption \citep[][]{maguire21}, i.e., they suggest that the source is not the accompanying CME. However, in these previous studies, an ongoing CME was observed at the same time by either The Large Angle Spectrographic Coronagraph \citep[LASCO;][]{br95} onboard the Solar and Heliospheric Observatory (SOHO) or by the coronagraphs of the Sun Earth Connection Coronal and Heliospheric Investigation \citep[SECCHI;][]{ho08} onboard the Solar Terrestrial Relations Observatory (STEREO) spacecraft, thus making the contribution of the CME shock to the type II emission unclear. A study that presents a type II observation in the absence of a CME suggested that it was generated by a flare blast wave \citep[][]{magdalenic12}, however, this study was carried out before the existence of the STEREO mission, which provided two additional coronagraphs in different heliocentric orbits monitoring the Sun. More recently, \citet{pankaj2016} presented a type II without a CME that was generated by an extreme ultraviolet (EUV) wave propagating as a fast shock interpreted in the study as being triggered by breakout reconnection along with the accompanying flare. A statistical study over two solar cycles \citep[][]{kumari2022} showed that the majority of type IIs were indeed associated with CMEs, with Solar Cycle (SC) 24 having a higher association with CMEs than SC 23, most likely due to a more comprehensive catalogue of CME detections by multiple spacecraft, LASCO and STEREO A and B. In SC 24, only a small percentage, 4.6\% of type IIs, had no CME association. Additionally, a study of moving radio bursts in SC24, mostly type II and type IV radio bursts, revealed that all moving bursts, but one, were closely associated with the occurrence of a CME \citep[][]{mo21}. The only type II burst not associated with a CME occurred on 25 October 2013 and was associated temporally to a weak C-class flare.  }

\begin{figure*}[ht]
\centering
    \includegraphics[width=0.88\linewidth]{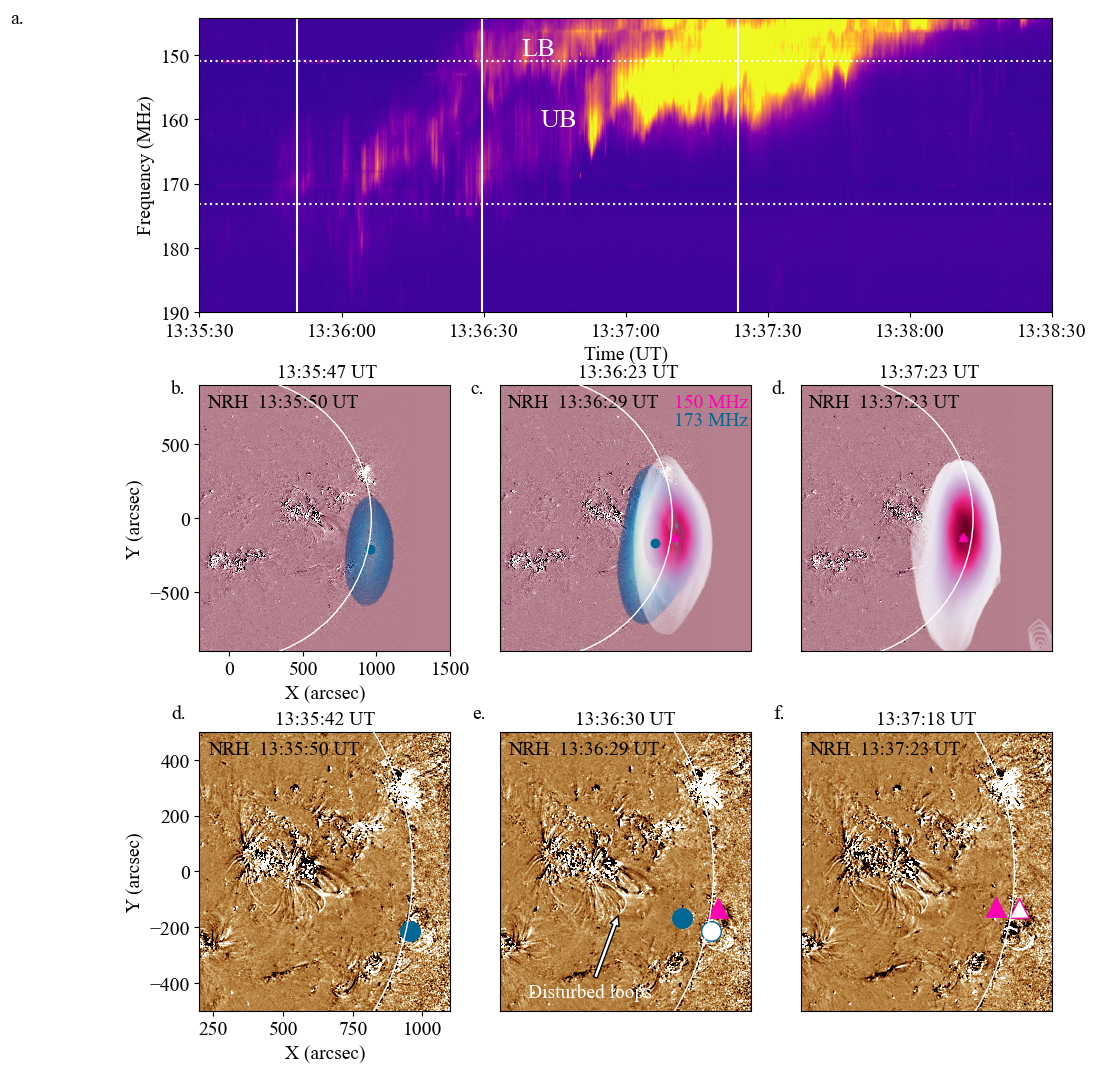}
    \caption{The locations of the split bands of the type II radio burst harmonic lane. (a) Zoomed-in dynamic spectrum showing the harmonic emission of the type II burst observed by ORFEES. The band-splitting is labelled as follows: lower frequency band (LB) and upper frequency band (UB). (b) AIA 211~\AA~ running difference images at three separate times showing the locations of the LB and UB bands at 150 (white to magenta filled contours) and 173~MHz (blue to green filled contours) corresponding to the vertical solid lines in (a). (c) AIA 211~\AA~ running difference images at three separate times overlaid with the centroids of the LB and UB bands of the corresponding radio sources in (b). The centroids are shown as a magenta triangle at 150~MHz and a blue circle at 173~MHz. The running difference images are two minutes apart. The white symbols in panel (e) represents the location of the UB at an earlier time in (d), while the white symbol in panel (f) reprsents the centroid of the LB at an earlier time in (e). }
    \label{fig:fig2}
\end{figure*}

{In this paper, we analyze in detail this type II radio burst and present the first detailed study that combines observations of a type II radio burst that occurs in the absence of a white-light CME with magneto-hydrodynamic simulations. In Sect.~\ref{sec:analysis}, we give an overview of the observations and data analysis techniques used. In Sect.~\ref{sec:results}, we present the results, which are further discussed in Sect.~\ref{sec:discussion}. The conclusions are presented in Sect.~\ref{sec:conclusion}.}

\begin{figure*}[ht]
\centering
    \includegraphics[width=0.8\linewidth]{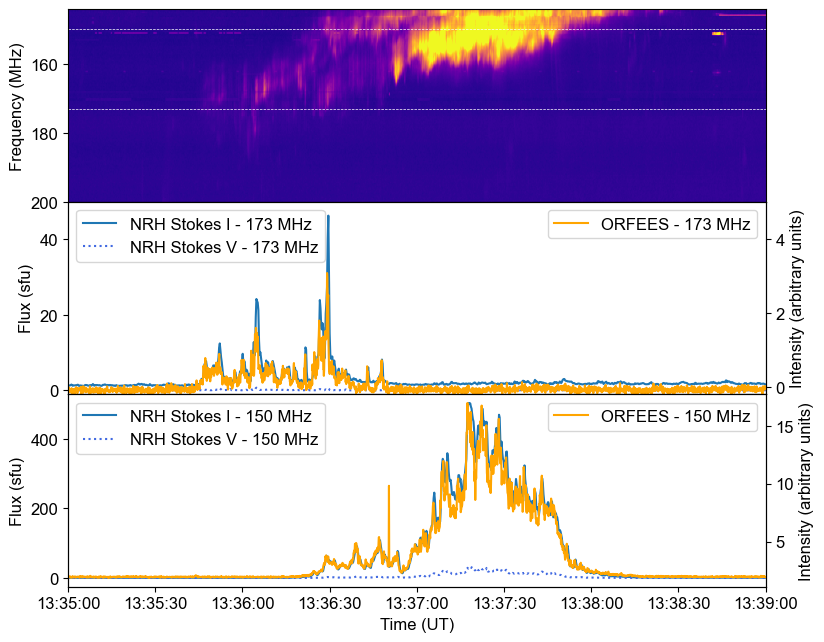}
    \caption{Zoomed-in dynamic spectrum of the type II burst and its flux density over time obtained from NRH images. The top panel shows the ORFEES dynamic spectrum during the type II harmonic lane. The last two panels show the flux of the type II bands together with the relative intensity extracted from the ORFEES dynamic spectrum at two frequencies: 173 (middle) and 150~MHz (bottom). These plots show that the radio sources imaged by the NRH represent the type II emission seen in the ORFEES dynamic spectrum.}
    \label{fig:fig3}
\end{figure*}


\section{Observations and data analysis} \label{sec:analysis}

\subsection{Radio emission}

{For a type II radio burst identified on 25 October 2013 no CME association could be made despite a good coverage of multi-view-point coronagraph observations \citep[][]{mo21}. No co-temporal CMEs were found in either LASCO C2 coronagraphs or the STEREO A and B COR1 and COR2 coronagraphs. The only eruptive signature that occurred during this Type II was the onset of a C2.3-class flare from the active region NOAA 11875 (Fig.~\ref{fig:fig1}a). The Type II radio burst was observed in dynamic spectra from the e-CALLISTO network \citep[][]{benz2005} and the ORFEES (Observation Radio pour FEDOME et l'{\'E}tude des {\'E}ruptions Solaires; \citealt[][]{Hamini2021}) radio spectrograph in the frequency range 50--190~MHz. The burst shows both fundamental and harmonic emission lanes with band-splitting and was preceded by a set of type III bursts (Fig.~\ref{fig:fig1}b). Some of the type III bursts did not continue to lower frequencies and show an inverted J-shape (labelled J-bursts in Fig.~\ref{fig:fig1}b). These J-bursts represent a sub-type of type III radio bursts where electrons do not escape along open magnetic field lines but instead propagate along closed coronal loops \citep[e.g.][]{re14,mo17}.}

{The type II burst lasts for $\sim$5~minutes and ends at a frequency of 50~MHz. At lower frequencies, there was no type II emission and only some interplanetary type III bursts were observed at the time. The harmonic lane of the type II burst is observed in images from the Nan{\c c}ay Radioheliograph \citep[NRH;][]{ke97} at 150 and 173~MHz with a cadence of 0.25~s. Both split bands of the harmonic lane can be imaged at these two frequencies. To determine movement of the Type II sources, we extracted the centroids of the Type II emission by fitting an eliptical Gaussian to the radio sources and estimating the errors in centroid positions using the methods of \citet{condon1997,mo19a}. The elliptical Gaussian was fitted to radio contours with values over $10^7$~K in brightness temperature which includes most of the extent of the radio sources observed. }

\subsection{Associated flare}

{A C2.3-class flare occurred at the same time as the type II and can be seen in the GOES X-ray time series in Fig.~\ref{fig:fig1}a in the yellow shaded region. The yellow region also denotes the extent of the dynamic spectrum in Fig.~\ref{fig:fig1}b. The flare is short-lived and peaks at 13:37~UT. The type III and type II bursts discussed above were observed during the rise and peak phase of the flare. The flare and the associated active region can also be observed in EUV images from the Atmospheric Imaging Assembly \citep[AIA;][]{le12} onboard the Solar Dynamics Observatory \citep[SDO;][]{pe12}. The type II radio sources are located on the western solar limb (for a full evolution of the radio bursts and the flare, see Movies~1 and 2 accompanying this paper). No coronal dimming was observed next to the active region, which would be a typical signature of a CME \citep[e.g.,][]{harrison2003}. }


\section{Results} \label{sec:results}

\begin{figure*}[ht]
\centering
    \includegraphics[width=0.95\linewidth]{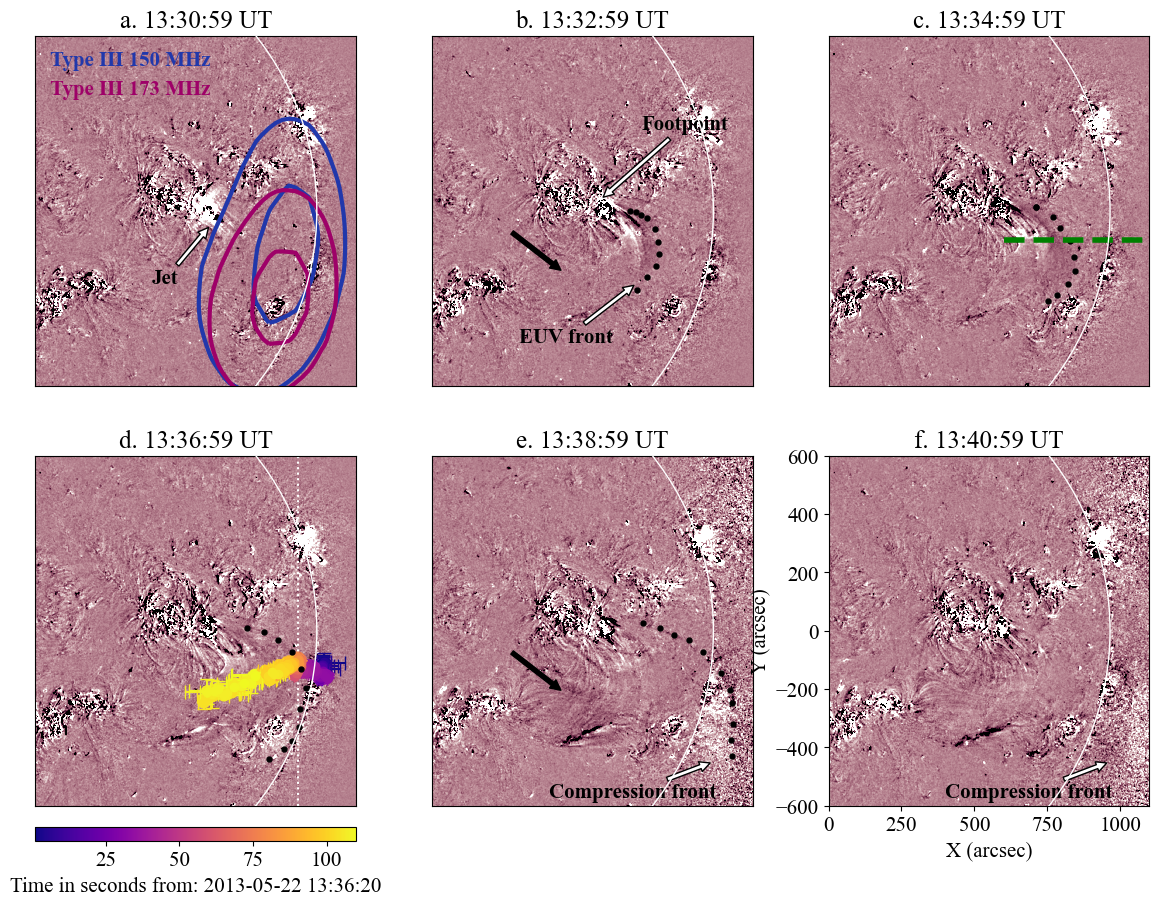}
    \caption{The propagation of the EUV front following the flare through the solar corona. AIA 211~\AA~ running difference images at six separate time showing the onset of the flare, followed by a EUV front propagating through the solar corona and faint moving feature detected beyond the solar limb. Type III radio bursts contours are overlaid in (a) at two frequencies: 150 (blue) and 173~MHz (magenta). The EUV front is labelled in panels b--e by the black dotted outline. The radio centroids of the type II burst at 150~MHz are overlaid in (d), where the colour bar represents time in seconds. The dotted vertical line in (d) also separates the UB (left of line) and LB (right of line) centroids. The red-coloured dashed line in (c) represents the location of the slit used in Fig.~\ref{fig:fig5}. }
    \label{fig:fig4}
\end{figure*}

\subsection{Location and characteristics of the type II burst }

{The type II burst studied here has the structure of a typical type II burst showing both fundamental and harmonic emission with split lanes. The split bands are labelled in the dynamic spectrum in Fig.~\ref{fig:fig2}a as follows: upper band (UB; higher-frequency band) and lower band (LB; lower-frequency band). One of the theories behind band-splitting in type IIs is that the upper and lower bands represent emission from downstream and upstream regions of the shock, respectively \citep{smerd1974}. However, this is not the case according to  other theories and some simulations that suggest they may originate at different regions upstream of the shock \citep[][]{mclean1967, knock2003}. }

\begin{figure}[ht]
\centering
    \includegraphics[width=0.95\linewidth]{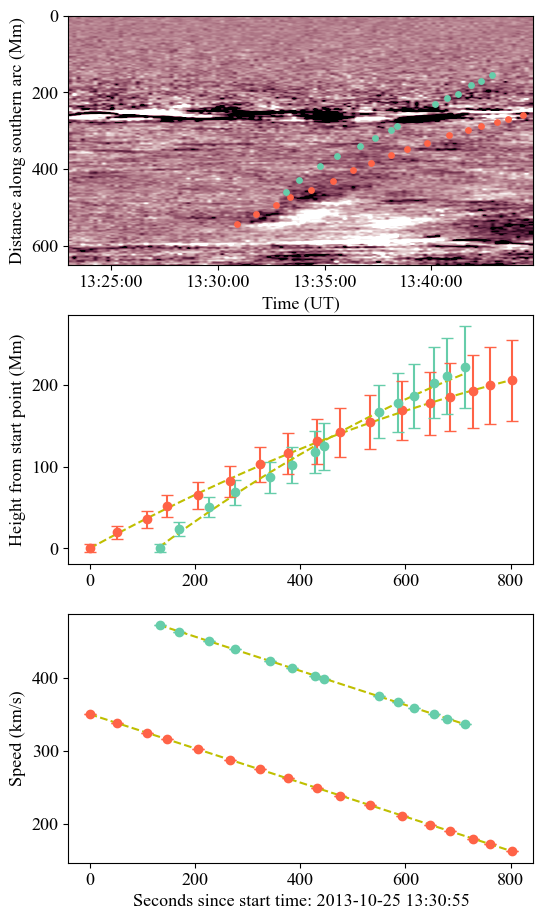}
    \caption{Propagation of the EUV front along the slit denoted by the red dashed line in Fig. 4c. (a) Emission inside the slit over time showing a propagation feature starting at $\sim$13:31~UT. (b) The height--time points of the propagating feature are plotted as orange circles. A fainter structure observed ahead of the propagating EUV front is plotted as green circles. (c) The speed of the propagating feature and fainter structure after fitting a height--time function to the data points in (b). }
    \label{fig:fig5}
\end{figure}

{What makes this burst unusual is that no co-spatial and co-temporal white-light CME association can be made (either in LASCO or STEREO coronagraphs) and that it is associated with a relatively weak C-class flare (C2.3). NRH images reveal a radio source at 150 and 173~MHz on the western solar limb above the active region NOAA 11875 at the same time as the type II burst (Fig.~\ref{fig:fig2}). The ORFEES dynamic spectrum is shown in a zoomed-in plot in Fig.~\ref{fig:fig2}a, together with filled contours of the radio sources at 150 and 173~MHz in Figs.~\ref{fig:fig2}b--c, and the centroids of these radio sources in Figs.~\ref{fig:fig2}d--f. The radio sources are overlaid on AIA 211~\AA~ running difference images of the Sun in Figs.~\ref{fig:fig2}b--c, while the centroids are overlaid on AIA 193~\AA~ running difference images of the Sun in Figs.~\ref{fig:fig2}d--f, at three separate times denoted by the solid white lines in Fig.~\ref{fig:fig2}a. Panels \textit{b} and \textit{d} show the radio source and centroid, respectively, at a time when the LB of the type II burst begins and can be imaged at 173 MHz. In panels \textit{c} and \textit{e}, we observe the time of both the UB (173~MHz) and the LB (150~MHz), while panels \textit{d} and \textit{f} only show the UB at 150~MHz, when the type II emission is most intense. There is a spatial difference between the source positions of the two split bands: the LB appears to be located further away from the Sun centre while the UB appears to be located closer to Sun centre. This separation is 96$\pm$9~Mm (133$\pm$12~arcseconds) at 13:36:29~UT (Figs.~\ref{fig:fig2}c and \ref{fig:fig2}e). Plane-of-sky projections effects may affect this separation in 3D and the de-projected separation is estimated in Section~3.3 following a 3D analysis and coronal modelling. There is also a small spatial separation between the bands at the same frequency but different times, however, this could be attributed to the movement of the type II burst in time.}


{The flux of the radio sources imaged by the NRH at 150 and 173~MHz is also compared to the relative intensity with the type II lanes in the ORFEES dynamic spectrum as shown in Fig.~\ref{fig:fig3}. In this figure, the flux in the Stokes I and Stokes V images is estimated by adding the pixels over the extent of the radio source that was defined as the pixels with values above 20\% of the maximum intensity level in each image. This flux is then compared with the relative intensity of the Type II obtained from the ORFEES dynamic spectrum. The ORFEES spectrum has a lower temporal resolution than the NRH images, 100~ms compared to 250~ms, respectively. Figs.~\ref{fig:fig3}b--c show the Stokes I (solid blue line) and Stokes V (dotted blue line) fluxes together with the ORFEES relative intensity of the type II at 173 and 150~MHz. The harmonic lanes imaged by the NRH are weakly polarised at 150~MHz (see the dashed blue line in Fig.~\ref{fig:fig3}c that represents the Stokes V flux). The degree of circular polarisation of this burst is negligible (<5\%). Fig.~\ref{fig:fig3} shows a good association between the imaged radio source and the relative intensity of the type II obtained from the dynamic spectrum. This is a confirmation that the imaged bursts represent the type II which shows good temporal and spatial association with the C-class flare.  }

{The radio centroids at 150~MHz over the full duration of the burst are shown in Figs.~\ref{fig:fig4}d. The colours of the centroids represent the time in seconds from the onset of the type II burst. The centroids show a movement back towards the flaring active region. Such movement has been observed before in the case of herringbones \citep[][]{mo19a, morosan2022}. The errors in the centroids locations are small ranging from 2 to 50~arcseconds and thus, the movement of the radio sources in the plane-of-sky (spanning over $\sim$200~arcseconds) is larger than the extent of the uncertainty in position. However, in that study the apparent movement towards the source active region was in fact a plane-of-sky projection effect and thus the centroids followed the three dimensional lateral expansion of the associated CME outside the plane-of-sky. In our case there is no CME present. Therefore, the shock wave responsible for the emission is likely to expand laterally across the disc in the low corona as a freely propagating shock expanding laterally, likely, in the shape of a dome \citep[e.g.,][]{hudson2003}. The centroids in Fig.~\ref{fig:fig4}d also include the time of occurrence of both split-bands, however, both split bands seem to travel in that direction. The apparent movement back towards the active can only be explained by the lateral expansion of the shock wave outside the plane-of-sky.}

\subsection{EUV and white-light signatures associated with the type II burst}

{At other wavelengths, any physical manifestation or indication of a CME and an associated shock wave appear to be absent. Before the occurrence of the type II burst, during the rise phase of the flare (between 13:30 and 13:34~UT), we observe a jet in the EUV images (labelled in Fig.~\ref{fig:fig4}a). During this phase we also see a number of type III radio bursts, where some of these are in fact J-bursts, indicating that some of the electron beams propagate along closed magnetic field lines. These Type IIIs are plotted as radio contours at 150 (blue) and 173~MHz (magenta) in Fig.~\ref{fig:fig4}a.}


\begin{figure*}[ht]
\centering
    \includegraphics[width=0.95\linewidth]{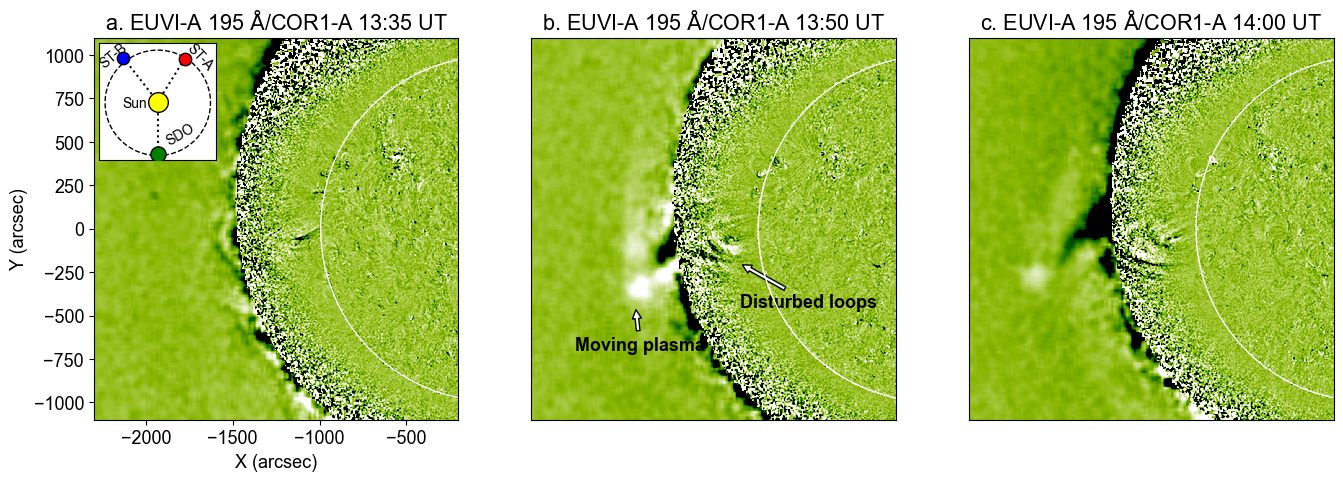}
    \caption{Disturbed loops and a moving feature observed beyond the solar limb in white-light images. STEREO-A white-light images from COR1 combined with EUVI 195~\AA~ images are shown at three consecutive times. The inset in panel (a) shows the locations of the STEREO spacecraft in relation to the Sun and SDO. }
    \label{fig:fig6}
\end{figure*}

{Shortly after the rise phase of the flare (after 13:34~UT), sets of disturbed loops appear (loops that have moved from their initial locations so that they become visible in running difference images) rooted with one footpoint in the flaring active region (see the black arrows in Figs.~\ref{fig:fig4}b and \ref{fig:fig4}e). The type II radio burst is located ahead of these disturbed loops in the plane-of-sky (see Fig.~\ref{fig:fig4}e). In the AIA 211~\AA{} channel, also in running difference images, a faint EUV front or wave is observed propagating towards the limb (Figs.~\ref{fig:fig4}b--f). This EUV front is indicated by the black dotted outline in the panels of Fig.~\ref{fig:fig4}. The propagating EUV front is likely the cause of the disturbed loops seen in the AIA 193~\AA{} images. After the EUV front reaches the solar limb, a faint signature of a propagating compression front is observed in (Figs.~\ref{fig:fig4}e--f), not, however, resembling a CME (the faint compression front is clearer in Movie 2 accompanying the paper as a faint movement beyond the solar limb). This propagating compression appears to move away from the solar limb in plane-of-sky images and it is the only signature of plasma moving beyond the solar disc in AIA images. This faint emission is visible in images at 193 and 211~\AA{} (please see the Movie~1 accompanying this paper to observe the movement of this compression front away from the solar limb). However, no white light signatures are visible after this time in the LASCO C2 coronagraph. }

{The speed of the propagating EUV front can be estimated by extracting the emission intensity over time inside a slit in AIA 211~\AA{} images \citep[e.g.,][]{kim2014,liu2018}. The location of the slit is along the dashed red line in Fig.~\ref{fig:fig4}c. The slit intensity over time shows a faint propagating feature that represents the movement of the EUV front across the solar disc (denoted by the green and red dotted lines in Fig.~\ref{fig:fig5}a) starting at $\sim$13:31~UT, which is $\sim$10 minutes after the onset of the flare at 13:10~UT. This propagating feature extends to the solar limb and beyond. This propagating feature can be viewed in two ways: a brighter outline denoted by the red dotted line and a fainter outline ahead of it, starting two minutes later at $\sim$13:33~UT, denoted by the green dotted line in Fig.~\ref{fig:fig5}a. The height--time profiles of both of these outlines are shown in Fig.~\ref{fig:fig5}b. The height--time plots are fitted with a function that contains a constant acceleration term:
\begin{equation}
    h = h_0 + v_0 t + \frac{1}{2}a_0 t^2 ,
\end{equation}
where $h_0, v_0$, and $a_0$ are the initial height, speed and acceleration, respectively. The results of the fit are shown in the speed--time plot in Fig.~\ref{fig:fig5}c.}

{The EUV front appears to be decelerating as it travels through the solar corona, however, initially it has a high speed of over 450~km/s. This speed is, however, a lower limit due to plane-of-sky projection effects, since the EUV wave propagates across a spherical surface. This speed is comparable to that of EUV waves that occur in close association with CMEs \citep[][]{long2008}. This indicates that the EUV front is initially an impulsive event most likely related to the flare and it decelerates as it propagates trough the corona as there is no longer a driving component present. At the time of the type II burst onset, the EUV front has a plane-of-sky speed of $\sim$280~km/s at the brighter outline and $\sim$430~km/s at the fainter outline. The fainter outline of the EUV front appears to be significantly faster with an initial speed of $\sim$470~km/s. At such speeds it is possible for a shock wave to form in the low solar corona if the driver propagates through an area of low Alfv\'en speed \citep[e.g.,][]{warmuth2005} which will be discussed in the next sub-section.} 

{The STEREO spacecraft provide additional vantage points to study this emission. The flare is observed as originating behind the limb in STEREO-A. Running difference images from STEREO-A EUVI and COR1 combined are shown in Fig.~\ref{fig:fig6} at three separate times. At 13:35~UT, a set of disturbed loops are visible outside the solar limb in EUVI 195~\AA~ images. After this time, COR1A shows a faint feature first appearing in its field-of-view at 13:50~UT that does not propagate further into the outer corona beyond $\sim$2.3~R$_\odot$ (for a complete evolution of this emission see Movie~3). This feature eventually disappears from the running difference images at 14:25~UT. The white-light feature likely represents the signature of either a failed eruption or bulk plasma motion following the jet or flare. This bulk plasma motion is likely to be connected to the EUV front observed from Earth's perspective.}

\begin{figure*}[ht]
\centering
    \includegraphics[angle = 0, trim = 0cm 0cm 0cm 0cm, width=15cm]{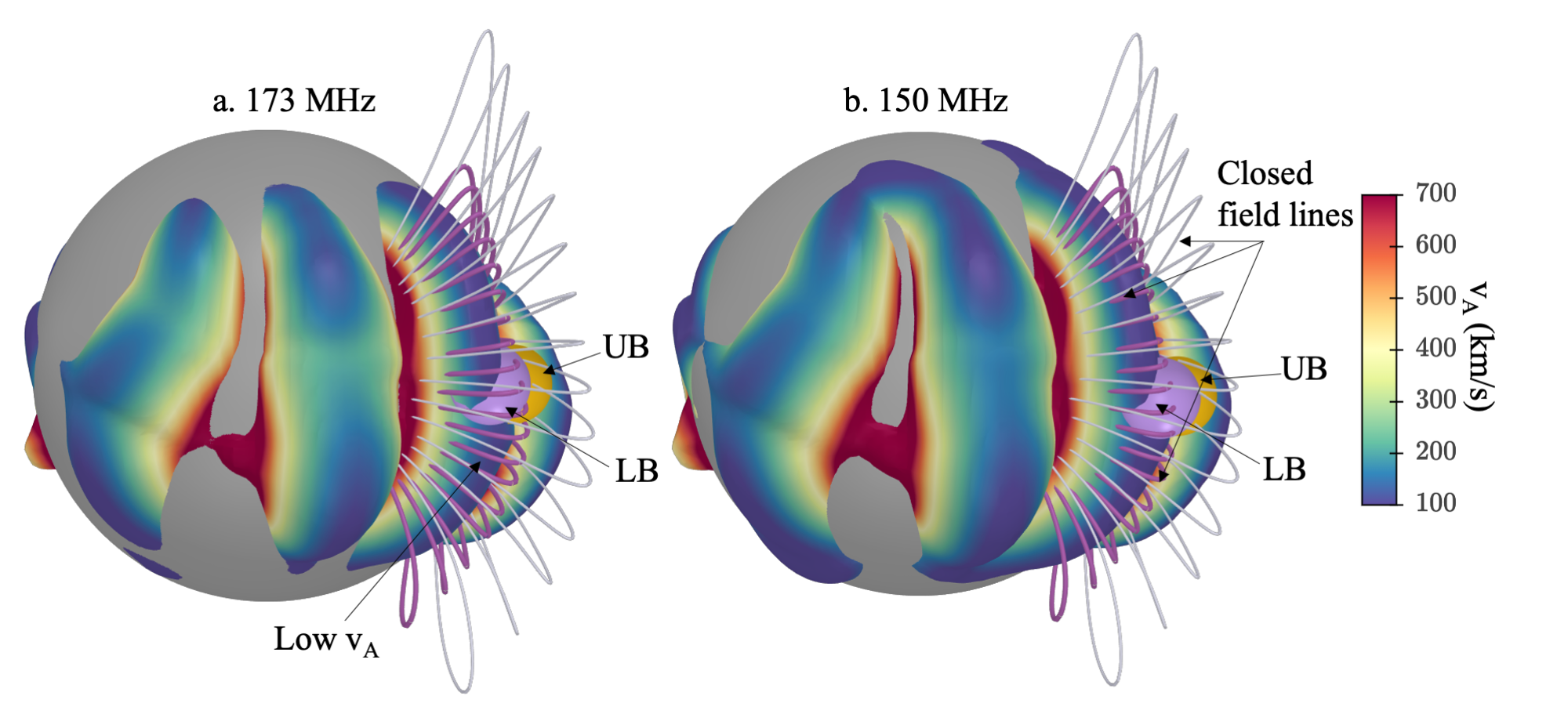}
    \caption{The macro-scale properties of the corona where the type II burst originates. The height of the plasma density level corresponding to the harmonic plasma frequency of 173 and 150~MHz is shown as a 3D surface in panels (a) and (b), respectively. The grey sphere has a radius of 1.2 R$_\odot$ to help visualize the enhanced density regions. Overlaid on the density surface are the values of Alfv\'en speed in km/s at the corresponding height. The 3D location of the UB and LB centroids of the type II burst are represented by orange and purple spheres, respectively, at the times presented in Figs.~\ref{fig:fig2}b--d. Also overlaid on the density surface in the region where the type II centroids originate are closed magnetic field lines forming a coronal streamer. }
    \label{fig:fig7}
\end{figure*}

\subsection{Coronal properties in the vicinity of the type II sources}

{To determine the properties of the plasma in which the type II radio sources propagate, we employed the Magnetohydrodynamics Around a Sphere Thermodynamic (MAST) model \citep{li09}. The MAST model is an MHD model developed by Predictive Sciences Inc.\footnote{\url{http://www.predsci.com/}} that can be used to study the large-scale structure and dynamics of the solar corona and inner heliosphere. The inner boundary conditions for the magnetic field in the model are photospheric magnetograms from the Heliospheric and Magnetic Imager \citep[][HMI;]{scherrer12} onboard SDO. The model also includes thermodynamic processes with energy equations accounting for thermal conduction, radiative losses, and coronal heating. This thermodynamic MHD model is capable of reproducing large-scale coronal features that are  observed in white light, EUV, and X-ray wavelengths \citep{li09}. The MAST model is limited by the input magnetogram, therefore, any features close to the solar limb at the time of the flare may not be accurately represented due to the obscured view of photospheric features near the limb. The MAST model is also not time-dependent, thus, any magnetic changes at the time of the flare are not taken into account, however, the MAST model produces a good representaiton of global coronal features such as the location of streamers.  }

{The MAST model outputs used in this study are global electron densities and magnetic field strengths. These properties are then used to compute the global Alfv\'en speed in the corona (Fig.~\ref{fig:fig7}) according to the methods of \citet[][]{morosan2022}. The results of the model are presented in Fig.~\ref{fig:fig7} for two cases: the plasma density height corresponding to harmonic plasma emission at 173~MHz (a) and the plasma density heights corresponding to harmonic plasma emission at 150~MHz (b). These density surfaces are obtained directly from the MAST model. The panels of Fig.~\ref{fig:fig7} show the global heights at which a certain level of electron plasma density occurs. The grey sphere denotes a height of 1.2~R$_\odot$ from the solar centre. The electron density is not uniform in the solar corona but instead shows ridges of increased density at certain locations. Overlaid on these density ridges, are values of the Alfv\'en speed where the colour map goes from blue (low Alfv\'en speed) to red (high Alfv\'en speed) and the values are in km/s (a colour bar is included to the right of Fig.~\ref{fig:fig7}).  }

\begin{figure*}[ht]
\centering
    \includegraphics[angle = 0, trim = 0cm 0cm 0cm 0cm, width=\linewidth]{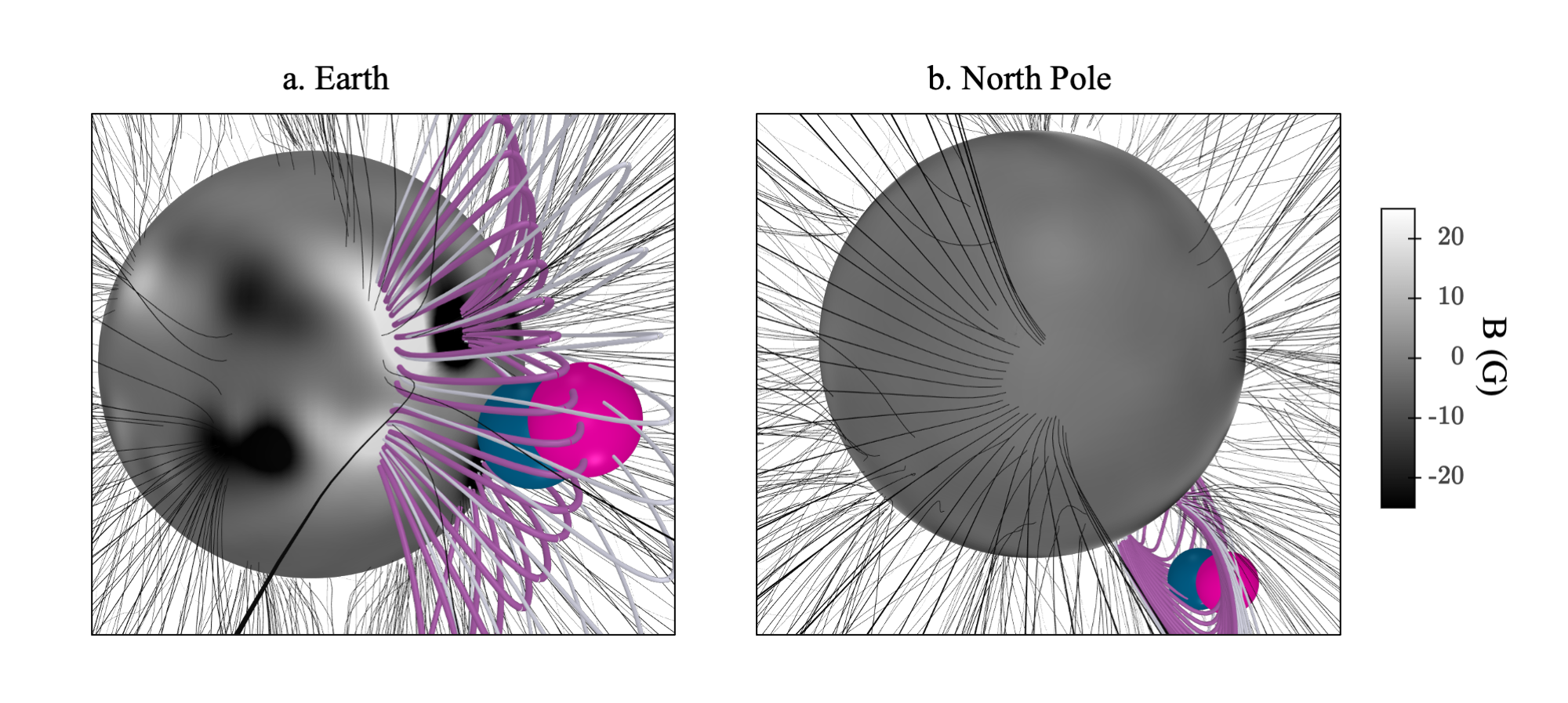}
    \caption{The 3D location of the splitbands at 13:36:29 UT. The 3D location of the UB (magenta sphere) and LB (blue sphere) at 13:36:29~UT together with closed (grey and magenta) and open (black) field lines obtained from the MAST model from the Earth perspective (a) and the solar North pole perspective (b). The UB and LB points correspond to the centroids in Fig.~\ref{fig:fig2}c.  }
    \label{fig:fig8}
\end{figure*}

{The centroids of the radio sources from Figs.~\ref{fig:fig2}d--f are overlaid on the model results at both 173 and 150~MHz. The purple spheres in Fig.~\ref{fig:fig7} represent the LB centroids at 13:35:42~UT for 173 MHz (a) and at 13:36:30~UT for 150 MHz (b). The orange spheres represent the UB centroids at 13:36:30~UT for 173~MHz (a) and at 13:37:18~UT for 150~MHz (b). These centroids are de-projected from the plane-of-sky view using the height of the electron density value corresponding to the appropriate plasma frequency at the location of the radio bursts. The de-projected height of the type II bands is 1.55~R$_\odot$ at 150~MHz (LB) and 1.47~R$_\odot$ at 173~MHz (UB). The x and y-coordinates of the centroids are matched with the appropriate height obtained from the MAST electron density model assuming harmonic plasma emission by converting their frequency to plasma density and extracting this density from the MAST density isosurfaces in Fig.~\ref{fig:fig7}). The spatial separation of the split bands can now be estimated using the de-projected centroids and it is 97~Mm (134~arcseconds). This value is very close to the plane-of-sky separation reported in the previous Section~3.1, indicating that plane-of-sky projection effects are small for the two different frequencies. The 3D separation of the split-bands at the same time can also be seen in Fig.~\ref{fig:fig8}. In this figure, the 3D location of the UB and LB at 13:36:29~UT are shown together with closed and open field lines obtained from the MAST model from two perspectives. }

{Also overlaid on the density ridges are coronal magnetic field lines obtained from the MAST model in the region where the type II sources originate. In this region, the magnetic field consists exclusively of closed field lines forming a coronal streamer. The multiple J-bursts associated with the flare that are labelled in Fig.~\ref{fig:fig1}a also occur close to the source locations of the type II burst (see Fig.~\ref{fig:fig4}a) and provide additional evidence to the existence of the closed field region where the type II propagates.}

{The type II centroids are located in a region with elevated densities inside the streamer. This enhanced density region is also in an area of minimum Alfv\'en speed ($\sim$100~km/s). A disturbance propagating at a speed of up to $\sim$430 ~km/s (such as the speed of the EUV front at the time of the Type II onset estimated from AIA images) is thus capable of driving a shock at these coronal heights. The type II burst is short lived and ends at 13:40~UT and a frequency of 50~MHz even though the EUV front is still propagating until $\sim$13:45~UT. This indicates that the type II emission does not propagate further out into the solar corona with the EUV front. This is likely due to the fact that as the EUV front propagates towards the edge of the visible solar disc, it then escapes the density ridge with low Alfv\'en speeds. Behind this ridge, there is a lower density region with Alfv\'en speed maxima, located close to the solar limb. In this region, characterised by Alfv\'en speeds of over 700~km/s, it is no longer possible for a decelerating disturbance to drive a shock. }


\section{Discussion} \label{sec:discussion}

{The only observational manifestations of a propagating structure associated with the type II burst observed in this study are a faint EUV front on the disc observed from Earth's perspective and plasma movement beyond the solar limb observed from STEREO A's perspective. The EUV front decelerates as it propagates through the solar corona. However, it is fast enough to drive a shock as it propagates inside a region of low Alfv\'en speed. The EUV front is presumably a low coronal manifestation of the shock that accelerates the type II producing electrons. The decelerating speed of the EUV front (however, only certain in the plane-of-sky) and the regions of high Alfv\'en speed following the closed field region indicates that the type II emission and possibly electron acceleration associated with this EUV front comes to a stop early during the flaring process. Thus, no type II emissions are seen beyond the peak phase of the flare, which agrees with the disturbance no longer steepening into a shock as it propagates further towards the limb. The type II also suddenly stops at a frequency of 50~MHz, coincident with the time when the EUV front approaches the edge of the visible solar disc (see Fig.~\ref{fig:fig5}), where a high Alfv\'en speed region begins based on the MAST model. }

{The EUV front consists of a fainter and faster feature (see the green dotted outline in Fig.~\ref{fig:fig5}a), which starts a few minutes after the onset of the brighter manifestation of the EUV front and reaches the edge of the solar disc before this brighter front. This feature may be interpreted as a signature of a shock wave accompanying the EUV front. The EUV wave appears to be a freely propagating pressure wave that steepens into a shock in the region of low Alfvén speed and generates a type II burst in that region.}

{In the absence of a clear CME signature, we present three possible scenarios that can lead to the generation of the observed type II burst: a failed eruption accompanied by a coronal wave that steepens into a shock wave, a jet from the flaring active region accompanied by a similar wave or the occurrence of a stealth CME driving a shock. }

{In white-light images from STEREO A, a small amount of plasma is observed moving beyond the solar limb following the flare. This motion is likely a signature of a failed eruption or a white-light signature of the solar jet that accompanies the eruption. A set of disturbed loops is observed at EUV wavelengths at lower heights but radially co-spatial with the plasma motion observed at larger heights. Simulations have shown that a failed eruption can be accompanied by bulk plasma motions that can launch a large-amplitude wave that can steepen into a shock and then propagate further out \citep[see for example Fig. 5.1 in][]{jens_thesis}. The EUV front observed in our analysis can thus be evidence of the large-amplitude wave launched by the bulk plasma motions accompanying the failed eruption.  A freely-propagating, refracting pressure wave has been observed before expanding in the lateral direction \citep[e.g.,][]{hudson2003}. This wave was tilted away from the vertical because of refraction towards regions of low Alfv\'en speed, since the Alfv\'en speed increases with height in the corona. In our case, if a freely-propagating coronal wave would be launched following the confined plasma motions in the flaring region, it would be refracting to the closed-field low Alfv\'en speed region it encounters, also in the region where the type II centroids are located and the region where the EUV front propagates and is likely to steepen into a shock wave. } 

{A jet is also present originating from the flaring active region during the rise time of the flare. Type III bursts occur at the same time as the jet and the faint EUV front is propagating across the disc following the direction of the material expelled by the jet (see for example Figs.~\ref{fig:fig4}a and b). The jet can also launch a freely propagating wave that is not only limited to the front of the jet but expands to the surrounding regions \citep[e.g.,][]{su2015,shen2018}. \citet[][]{su2015} present a similar scenario in which a Type II radio burst was observed without being associated with a CME but co-temporal with a jet. However, they lacked radio imaging observations to determine that the type II burst was indeed associated with the observed EUV wave. }

{Another possibility is the occurrence of a stealth CME \citep[e.g.][]{ly16, nitta21} which would require the analysis of in situ data to confirm. However, there was a faster halo CME observed on the same date after the event analysed in this paper with a first detection by LASCO C2 at 15:12~UT. The faster halo CME originated from a different active region and it was accompanied by its own radio emission detailed in \citet[][]{morosan2022}. This halo CME would have likely merged or interacted with an earlier stealth CME, if launched, and thus the in situ data would likely not show any clear CME signatures other than the fast halo CME. This was confirmed by investigating the magnetic field and plasma data from STEREO-A and spacecraft at Earth (which are the most likely to have been impacted by a possible stealth CME associated with this event). A stealth CME cannot, however, be fully excluded as the driver of the shock wave due to the fact that some material is observed to move away from the solar limb in AIA and COR1-A images. The Type II centroids also show apparent movement towards the source active region. Such a motion towards the source region is likely to be a projection effect in the plane of sky as the radio burst occurs at larger heights than the active region observed in EUV. This motion can be explained by some form of lateral expansion of the shock wave in three dimensions \citep[e.g.,][]{mo19a}. Such lateral expansions are not limited to CME-driven shocks \citep[e.g.,][]{morosan2022, mo19a}, but can also occur in freely propagating large-amplitude MHD waves. Previous observations of coronal waves observed in soft X-rays also associated with metric Type II bursts were interpreted as freely propagating, laterally expanding shocks \citep[][]{hudson2003}.  }

{Our analysis shows that there is a clear spatial separation between the two bands both at the same time and different frequencies and at the same frequency and different times (Fig.~\ref{fig:fig2}). A spatial separation between the split bands at the same frequency but at different times has been reported before \citep[e.g.,][]{Smerd1975}. However, both in \citet[][]{Smerd1975} and the present study, the type II burst is likely to move in position over a time period of $\sim$1~minute at the same frequency. This is also evident in our study since each individual band moves through time as shown in Fig.~\ref{fig:fig4}d and it represents a limitation to estimating the separation at the same frequency. In our study, there is also a clear spatial separation between the bands at the same time but different frequencies. \citet{ho83} suggest a separation of $\sim$0.1~R$_\odot$ ($\sim$70~Mm) that corresponds to a typical splitting of 18 MHz for the harmonic lane at 80 MHz. Our split-band separation is: 96~Mm (0.14~R$_\odot$), which is in agreement to the separation estimated by \citet{ho83} that used a simple electron density model to estimate heights from the emission frequency \citep{new61}. Our results are, however, in disagreement to the theory suggested by \citet[][]{Smerd1975} and the more recent study of \citet{chrysaphi2018} that found that there is no separation between the UB and LB centroids at the same time but different frequencies. A more recent study with the Murchinson Widefield Array \citep{bhunia2023} has also found a significant separation between the UB and LB bands in agreement with our study and the theory suggested by \citet{mclean1967} and \citet{ho83} that band-splitting is likely explained by the two bands originating from spatially distinct locations upstream of the shock and, therefore, not in adjacent upstream/downstream regions.}





\section{Conclusion} \label{sec:conclusion}

{We presented observations of a type II radio burst that occurs in the absence of a CME detection. The type II burst is most likely caused by a freely propagating pressure wave that manifests itself as an EUV front propagating away from the flaring active region and steepens to a fast-mode shock-wave in a region with low Alfv\'en speeds. As indicated in previous studies, we suggest that a CME eruption is not always required to drive a shock wave to produce type II radio bursts in the solar corona using radio imaging observations. Instead, a wave propagating through a low Alfv\'en speed region with a speed of $\sim$400~km/s is sufficient to become a shock that then generate a type II radio burst. }


{We also showed that the type II split bands have different spatial locations, which are similar to other type IIs observed related to CME shocks \citep[e.g.,][]{bhunia2023}. Future high-spatial resolution observations are necessary to study the band-splitting phenomena in type II bursts to determine the spatial locations of the two bands. These observations are possible with the Low Frequency Array (LOFAR) or the Murchinson Widefield Array (MWA). In particular, the higher frequency channels of these arrays can be used to study the harmonic emission of type II lanes since the harmonic is less likely to be affected by radio propagation effects in the corona. }


\begin{acknowledgements}{D.E.M. acknowledges the Academy of Finland project `RadioCME' (grant number 333859). D.E.M. and. A.K. acknowledge the University of Helsinki Three Year Grant. J.P. acknowledges the Academy of Finland Project 343581. E.K.J.K. acknowledges the ERC under the European Union's Horizon 2020 Research and Innovation Programme Project SolMAG 724391, and Academy of Finland Project 310445. All authors acknowledge the Finnish Centre of Excellence in Research of Sustainable Space (Academy of Finland grant numbers 312390, 312357, 312351 and 336809). We thank the Radio Solar Database service at LESIA \& USN (Observatoire de Paris) for making the NRH/ORFEES data available. We also thank the eCALLISTO network for the continuous availability of radio spectra.} \end{acknowledgements}

\bibliographystyle{aa} 
\bibliography{Bib_AA.bib} 

\end{document}